\documentclass[]{pasj01}

\usepackage{url}

\begin{document} 
\Received{}
\Accepted{}

\title{Systematic Identification of LAEs for Visible Exploration and 
Reionization Research Using Subaru HSC (SILVERRUSH). I. \\
Program Strategy and Clustering Properties of\\
$\sim 2,000$ Ly$\alpha$ Emitters at $z=6-7$\\
over the $0.3-0.5$ Gpc$^2$ Survey Area
}



\author{
	Masami \textsc{Ouchi},\altaffilmark{1,2}
        Yuichi \textsc{Harikane},\altaffilmark{1,3}
        Takatoshi \textsc{Shibuya},\altaffilmark{1}
        Kazuhiro \textsc{Shimasaku},\altaffilmark{4,5}
        Yoshiaki \textsc{Taniguchi},\altaffilmark{6}
        Akira \textsc{Konno},\altaffilmark{1,4}
        Masakazu \textsc{Kobayashi},\altaffilmark{7}
        Masaru \textsc{Kajisawa},\altaffilmark{8}
        Tohru \textsc{Nagao},\altaffilmark{8}   
        Yoshiaki \textsc{Ono},\altaffilmark{1}
        Akio K. \textsc{Inoue},\altaffilmark{9}
        Masayuki \textsc{Umemura},\altaffilmark{10}
        Masao \textsc{Mori},\altaffilmark{10}
        Kenji \textsc{Hasegawa},\altaffilmark{11}
        Ryo \textsc{Higuchi},\altaffilmark{1} 
        Yutaka \textsc{Komiyama}\altaffilmark{12,13}
        Yuichi \textsc{Matsuda},\altaffilmark{12,13}
        Kimihiko \textsc{Nakajima},\altaffilmark{14}
        Tomoki \textsc{Saito},\altaffilmark{15}  and
        Shiang-Yu \textsc{Wang},\altaffilmark{16} 
        %
        }
\altaffiltext{1}{Institute for Cosmic Ray Research, The University of Tokyo, 5-1-5 Kashiwanoha, Kashiwa, Chiba 277-8582, Japan}
\email{ouchims@icrr.u-tokyo.ac.jp}
\altaffiltext{2}{Kavli Institute for the Physics and Mathematics of the Universe (Kavli IPMU, WPI), The University of Tokyo, 5-1-5 Kashiwanoha, Kashiwa, Chiba, 277-8583, Japan}
\altaffiltext{3}{Department of Physics, Graduate School of Science, The University of Tokyo, 7-3-1 Hongo, Bunkyo, Tokyo, 113-0033, Japan}
\altaffiltext{4}{Department of Astronomy, Graduate School of Science, The University of Tokyo, 7-3-1 Hongo, Bunkyo, Tokyo 113-0033, Japan}
\altaffiltext{5}{Research Center for the Early Universe, Graduate School of Science, The University of Tokyo, 7-3-1 Hongo, Bunkyo, Tokyo 113-0033, Japan}
\altaffiltext{6}{The Open University of Japan, Wakaba 2-11, Mihama-ku, Chiba 261-8586, Japan}
\altaffiltext{7}{Faculty of Natural Sciences, National Institute of Technology, Kure College, 2-2-11 Agaminami, Kure, Hiroshima 737-8506, Japan}
\altaffiltext{8}{Research Center for Space and Cosmic Evolution, Ehime University, Bunkyo-cho, Matsuyama, Ehime 790-8577, Japan}
\altaffiltext{9}{Department of Environmental Science and Technology, Faculty of Design Technology, Osaka Sangyo University, 3-1-1, Nagaito, Daito, Osaka 574-8530, Japan}
\altaffiltext{10}{Center for Computational Sciences, University of Tsukuba, 1-1-1 Tennodai, Tsukuba 305-8577 Ibaraki, Japan}
\altaffiltext{11}{Department of Physics and Astrophysics, Nagoya University Furo-cho, Chikusa-ku, Nagoya, Aichi 464-8602, Japan}
\altaffiltext{12}{National Astronomical Observatory of Japan, 2-21-1 Osawa, Mitaka, Tokyo 181-8588}
\altaffiltext{13}{The Graduate University for Advanced Studies (SOKENDAI), 2-21-1 Osawa, Mitaka, Tokyo 181-8588}
\altaffiltext{14}{European Southern Observatory, Karl-Schwarzschild-Str. 2, D-85748 Garching bei Munchen, Germany}
\altaffiltext{15}{Nishi-Harima Astronomical Observatory, Center for Astronomy, University of Hyogo, 407-2 Nishigaichi, Sayo-cho, Sayo, Hyogo 679-5313, Japan}
\altaffiltext{16}{Academia Sinica, Institute of Astronomy and Astrophysics 11F of AS/NTU Astronomy-Mathematics Building, No.1, Sec. 4, Roosevelt Rd, Taipei 10617, Taiwan}

\altaffiltext{\ddag}{Based on data obtained with the Subaru Telescope.
The Subaru Telescope is operated by the National Astronomical Observatory of Japan.
}




\maketitle

\begin{abstract}
We present the SILVERRUSH program strategy and 
clustering properties investigated with
%
$\sim 2,000$
%
%
Ly$\alpha$ emitters at $z=5.7$ and $6.6$
found in the early data of the Hyper Suprime-Cam (HSC)
Subaru Strategic Program survey exploiting the carefully designed narrowband filters.
We derive angular correlation functions with
the unprecedentedly large samples of LAEs at $z=6-7$ over 
the large total area of $14-21$ deg$^2$ corresponding to $0.3-0.5$ comoving Gpc$^2$.
We obtain the average large-scale bias values of $b_{\rm avg}=4.1\pm 0.2$ ($4.5\pm 0.6$)
at $z=5.7$ ($z=6.6$) for $\gtrsim L^*$ LAEs, indicating
the weak evolution of LAE clustering from $z=5.7$ to $6.6$. 
We compare the LAE clustering results with
two independent theoretical models that suggest
an increase of an LAE clustering signal by the 
patchy ionized bubbles at the epoch of reionization (EoR),
and estimate the neutral hydrogen fraction 
%
%
%
to be $x_{\rm HI}=0.15^{+0.15}_{-0.15}$ 
%
%
at $z=6.6$. 
Based on the halo occupation distribution models, we find that
the $\gtrsim L^*$ LAEs are hosted by the dark-matter halos
with the average mass of $\log (\left < M_{\rm h} \right >/M_\odot) =11.1^{+0.2}_{-0.4}$
($10.8^{+0.3}_{-0.5}$) at $z=5.7$ ($6.6$) with a Ly$\alpha$ duty cycle of 1 \% or less,
where the results of $z=6.6$ LAEs may be slightly biased, due to the increase of 
the clustering signal at the EoR.
Our clustering analysis reveals the low-mass nature
of $\gtrsim L^*$ LAEs at $z=6-7$, and that these 
LAEs probably evolve into massive super-$L^*$ galaxies
in the present-day universe.

\end{abstract}

\section{Introduction}
\label{sec:introduction}

%

Ly$\alpha$ emitters (LAEs) are star-forming galaxies (and AGNs) with a strong Ly$\alpha$ emission line.
In the 1960s, it 
%
%
was
%
%
theoretically predicted that such galaxies are candidates of very young galaxies 
residing at $z\sim 10-30$ \citep{partridge1967}. About 30 years after the predictions, 
deep observations identified a few LAEs around AGN regions at $z=4.6$ \citep{hu1996}
and $z=2.4$ \citep{pascarelle1996}. Subsequently, LAEs in blank fields are
routinely found by deep and wide-field narrowband observations
conducted under Hawaii Survey \citep{cowie1998}, Large Area Lyman Alpha
Survey \citep{rhoads2000}, and Subaru Surveys (e.g. \cite{ouchi2003}).
These blank-field LAE surveys reveal various statistical properties of LAEs up to
$z\sim 7$ including the evolution of Ly$\alpha$ luminosity functions (LFs;
\cite{malhotra2004,kashikawa2006,ouchi2008,deharveng2008,cowie2010,
ouchi2010,hu2010,kashikawa2011,konno2014,matthee2015,santos2016,zhengzhen-ya2017,ota2017}).
%
%
Moreover, multi-wavelength follow-up imaging and spectroscopy for LAEs reveal
that the typical ($\sim L^*$) LAEs ($L_{\rm Ly\alpha}=10^{42}-10^{43}$ erg s$^{-1}$) at $z\sim 2-3$ 
have a stellar mass of $10^8-10^{9}$ M$_\odot$ \citep{gawiser2007,hagen2014},
a star-formation rate (SFR) of $\sim 10$ M$_\odot$ \citep{nakajima2012},
and a gas-phase metallicity of $\sim 0.1Z_\odot$ \citep{finkelstein2011,guaita2013,kojima2016}.
LAEs have strong high-ionization lines such as {\sc [Oiii]}5007, indicative of
the high ionization state of the inter-stellar medium (ISM) given by intense ionizing radiation
from young massive stars \citep{nakajima2014}. Outside the highly-ionized ISM, 
LAEs show diffuse extended Ly$\alpha$ halos in the circum-galactic medium (CGM) 
up to a few 10 kpc \citep{hayashino2004,steidel2011}.
%
%
The bright Ly$\alpha$ halos extending over a few hundred kpc
are known as Ly$\alpha$ blobs whose total Ly$\alpha$ luminosities are $\sim 10^{44}$ erg s$^{-1}$
(\cite{steidel2000,matsuda2004}, cf. \cite{saito2006}),
more than an order of magnitude brighter than the diffuse Ly$\alpha$ halos \citep{momose2014,momose2016}.
Deep spectroscopic observations identify large spatial overdensities of LAEs 
up to $z\sim 6$ \citep{shimasaku2003,ouchi2005a} that are possibly progenitors
of massive galaxy clusters found today. Recent deep spectroscopy also detect strong Ly$\alpha$ emission of $\sim 10$ LAEs
at $z=7.0-8.7$, pushing the redshift frontier of LAEs, to date (e.g. \cite{pentericci2011,ono2012,
shibuya2012,finkelstein2013,schenker2014,oesch2015,zitrin2015}).


There are three major scientific goals in recent LAE studies. 
The first goal is characterizing the nature of low-mass young 
galaxies at high-$z$. Because LAEs are low stellar mass galaxies with strong Ly$\alpha$
indicative of a starburst with young massive stars producing a large amount of ionizing photons 
\citep{ouchi2013,sobral2015,stark2015}, LAEs are used as probes of young galaxies. 
This motivation is the same as the one of the original predictions of LAEs given by \citet{partridge1967}. 
The second goal is to understand the density and dynamics of {\sc Hi} clouds in the ISM
and the CGM of star-forming galaxies. Due to the resonance nature of Ly$\alpha$ emission,
Ly$\alpha$ photons are scattered by {\sc Hi} gas in the ISM and the CGM,
and the dynamical properties of Ly$\alpha$ sources are not simply understood with
the observed Ly$\alpha$ velocity fields. Instead, both the density and kinematics of 
{\sc Hi} gas in the ISM and the CGM are encoded 
in the observed Ly$\alpha$ line velocity and spatial profiles. 
The combination of the Ly$\alpha$ line profile observations and models 
provide important information about the {\sc Hi} gas of the ISM and 
the CGM \citep{verhamme2008,zheng2011,shibuya2014,hashimoto2015,momose2016}.
The third goal is to reveal the cosmic reionization history and properties.
Due to the strong damping absorption of Ly$\alpha$ by the IGM 
{\sc Hi} gas at the epoch of reionization (EoR; $z>6$), Ly$\alpha$ emission of LAEs
are absorbed in the partially neutral IGM. The Ly$\alpha$ absorption 
depress the observed Ly$\alpha$ luminosities of LAEs, which make a decrease
of the Ly$\alpha$ LF towards the early stage of the EoR 
\citep{malhotra2004,kashikawa2006,ouchi2010,hu2010,kashikawa2011,santos2016,ota2017}.
Moreover, the clustering signal of observed LAEs
is boosted by the existence of the ionized bubbles \citep{furlanetto2006,mcquinn2007,ouchi2010},
because the Ly$\alpha$ absorption is selectively weak for LAEs residing in the ionized bubbles of the IGM at the EoR.
The Ly$\alpha$ LFs and clustering of LAEs are important quantities to
characterize the cosmic reionization history and ionized bubble topologies at the EoR.

%
%

There are three new large instruments that are used for LAE observations;
VLT/Multi-Unit Spectroscopic Explorer (MUSE; \cite{bacon2010}),
Hobby-Eberly Telescope Dark Energy Experiment/Visible Integral-Field Replicable Unit Spectrograph (HETDEX/VIRUS; \cite{hill2012}),
and
Subaru/Hyper Suprime-Cam (HSC; \cite{miyazaki2017}).
These three instruments can cover the complementary parameter space of LAEs
in redshift, depth, and survey volume.
Specifically, Subaru/HSC has a capability to take large-area ($>10$ deg$^2$) deep images
with the custom-made narrowbands (Section \ref{sec:strategy}) targeting LAEs at $z\sim 2-7$.
The combination of Subaru/HSC imaging and deep follow-up spectroscopy
allows us to address the key issues of LAEs for accomplishing the three major goals described above.

%

%


One of the most unique studies realized with Subaru/HSC is clustering of LAEs at the EoR (i.e. $z\gtrsim 6$)
that require high-sensitivity observations over a large area of the sky,
%
%
as we show in this study.
%
%
LAE clustering signals depend
on the hosting dark-matter halo properties of LAEs, and the distribution of the patchy ionized IGM (i.e. ionized bubbles)
that allows Ly$\alpha$ photons to escape from the partially neutral IGM at the EoR \citep{furlanetto2006,mcquinn2007,ouchi2010}.
%
%
%
Theoretical models suggest that a Ly$\alpha$ LF measurement, a popular statistical quantity of LAEs, has a degeneracy
between the Ly$\alpha$ escape fraction $f_{\rm esc}^{\rm Ly \alpha}$ 
(a flux ratio of observed to intrinsic Ly$\alpha$ emission; \cite{ledelliou2006,mao2007,kobayashi2010})
and duty cycle $f_{\rm duty}^{\rm LAE}$ (an abundance ratio of Ly$\alpha$ emitting galaxies to dark-matter haloes;
\cite{nagamine2010}).
%
%
Moreover, at the EoR, there is another degeneracy between the ionized bubble topology and neutral 
hydrogen fraction $x_{\rm HI}$ of the IGM \citep{kakiichi2016}. Theoretical studies show that 
these degeneracies can be resolved with the combination of Ly$\alpha$ LF
and clustering measurements 
%
%
(e.g. \cite{nagamine2010,sobacchi2015,hutter2015,kakiichi2016}).
%
%
The LAE samples at the post EoR, e.g. $z=5.7$, (the EoR, e.g. $z=6.6$)
are useful to determine $f_{\rm esc}^{\rm Ly \alpha}$ and $f_{\rm duty}^{\rm LAE}$.
%
%
%

%

In this paper, we describe our narrowband filters and the program strategy of the HSC LAE studies
named 'Systematic Identification of LAEs for Visible Exploration and Reionization Research 
Using Subaru HSC' (SILVERRUSH; Section \ref{sec:strategy}), and 
present our LAE samples (Section \ref{sec:sample}).
We obtain measurements of LAE clustering (Section \ref{sec:results}) that are calculated with 
the early samples of HSC LAEs at $z=5.7$ and $6.6$ \citep{shibuya2017a} based on
the Hyper Suprime-Cam Subaru Strategic Program (SSP; HSC et al.) data whose first data release is
presented in \citet{aihara2017}. Based on the LAE clustering measurements, we discuss 
cosmic reionization history and dark-matter halo properties of LAEs by the comparison with
the $\Lambda$CDM structure formation models in Section \ref{sec:discussion}.
Throughout this paper, magnitudes are in the AB system. We adopt a cosmology
parameter set of ($h$, $\Omega_{\rm m}$,  $\Omega_{\rm \Lambda}$, $n_s$, $\sigma_8$)
$=(0.7, 0.3, 0.7, 1.0, 0.8)$ consistent with the latest Planck results \citep{planck2016}.

\section{Narrowband Filters and the Program Strategy for the LAE Studies}
\label{sec:strategy}

In our LAE studies, we use four custom-made narrowband
filters that are fabricated for the HSC SSP survey.
We design the HSC narrowband filters targeting the OH sky windows
at 816, 921, and 1010 nm that correspond to $NB816$ (PI: Y. Taniguchi), 
$NB921$ (PI: M. Ouchi), and $NB101$ (PI: K. Shimasaku)
filters identifying the redshifted Ly$\alpha$ of LAEs at $z=5.7$, $6.6$, and $7.3$, 
respectively. These three narrowband filters densely cover LAEs from the 
%
%
heart
%
%
of the EoR 
to the post-EoR. Another HSC narrowband filter is $NB387$ (PI: K. Shimasaku) 
whose central wavelength is 387 nm. The $NB387$ band is the bluest
narrowband among the existing HSC filters, targeting the lowest redshift LAEs identified by 
the HSC observations.
The central wavelength of 387 nm is specifically chosen, because this band identifies LAEs at $z=2.2$
%
%
%
whose major strong lines of {\sc [Oii]3727}
and H$\alpha$6563 fall in the OH sky windows of $1.19$ and $2.10\mu$m,
respectively. The other major strong lines of H$\beta$4861 and {\sc [Oiii]5007} of $z=2.2$ LAEs are placed
at the $H$ band, where the effects of 
the OH sky emission are unavoidable, 
due to no clear OH windows.
%
%

We carefully determine the full width at half maximum (FWHM) value for each narrowband filter transmission
curve. First, we determine the FWHM of $NB921$ that properly covers a clear OH window
at the reddest band, allowing the typical transmission curve errors of the filter fabrication
\footnote{
There are some more OH windows redder than 921 nm, which include the 1010-nm window for the $NB101$ filter.
However, these OH windows in the red band are contaminated by moderately strong OH lines, unlike the one of 921 nm.
}.
%
%
%
We aim to accomplish the same detection limit of the rest-frame
Ly$\alpha$ equivalent width $EW_0$ in our samples of 
LAEs at $z=2.2$, $5.7$, and $6.6$.
An observed-frame Ly$\alpha$ equivalent width $EW_{\rm obs}$
has the relation of $EW_{\rm obs} = (1+z) EW_0$.
Because a narrowband flux excess is proportional (inversely-proportional) to
$EW_{\rm obs}$ (a narrowband FWHM value),
we scale the FWHM value of the $NB921$ band ($FWHM_{\rm NB921}$)
by $(1+z)$ to design the FWHM values of $NB387$ ($FWHM_{\rm NB387}$) 
and $NB816$ ($FWHM_{\rm NB816}$). 
More specifically, the designed FWHM values keep the relations of
$FWHM_{\rm NB387}=FWHM_{\rm NB921} (1+2.18)/(1+6.58)$
and
$FWHM_{\rm NB816}=FWHM_{\rm NB921} (1+5.71)/(1+6.58)$,
where the numbers of 2.18, 5.71, and 6.58 correspond to the redshifts of the
Ly$\alpha$ lines ($121.567$nm) falling 
in the centers of the $NB387$, $NB816$, and $NB921$ bandpasses, respectively.
%
%
Here, the FWHM of the $NB101$ band does not follow this scaling relation.
Instead, we choose an FWHM that is the narrowest limit
that can be reasonably accomplished by the present technology of the HSC narrowband filter fabrication.
This is because Ly$\alpha$ lines of the $NB101$ LAEs fall in the wavelength range
where the HSC imaging cannot reach a very deep magnitude limit, due to 
the relatively low CCD quantum efficiency and the unclean OH window.
In this way, we obtain the designed specifications of 
%
(central wavelength, FWHM wavelength) = 
(387.0nm, 5.5nm) for $NB387$,
(816.0nm, 11.6nm) for $NB816$,
(921.0nm, 13.1nm) for $NB921$, and
(1009.5nm, 9.0nm) for $NB101$.
%
%

\begin{figure}
 \begin{center}
  \includegraphics[width=8cm]{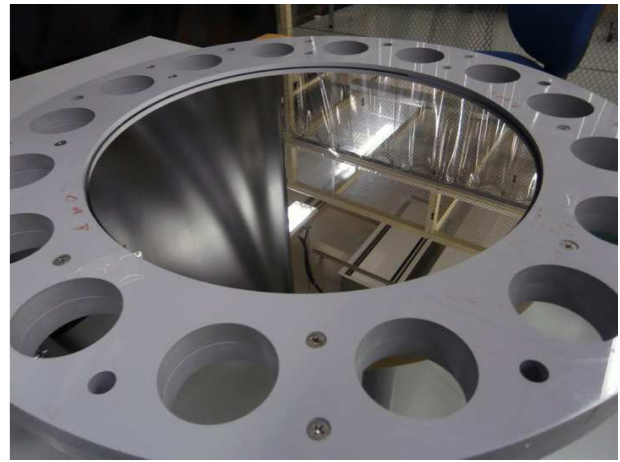} 
\end{center}
\caption{
HSC $NB921$ filter in a filter holder.
\label{fig:hsc_nb921}}
\end{figure}

The HSC narrowband filters are made of a B270 or quarts glass with multi-layer interferometric 
coatings that make the sharp cut-on and cut-off bandpasses for the narrowband filters.
The diameter of the HSC narrowband filter is large, 600 mm. Although 
uniform coatings are key to accomplish the spatially-homogeneous 
transmission of the narrowband filters, small non-uniformities are included in
the fabrication processes, which give the filter characterization slightly 
different from the design. Nevertheless, we have tried to make the coating distribution 
as uniformly as possible to achieve spatially-homogeneous transmission curves
for the narrowband filters. The picture of $NB921$ is shown in Figure \ref{fig:hsc_nb921}.

\begin{figure}
 \begin{center}
  \includegraphics[width=8cm]{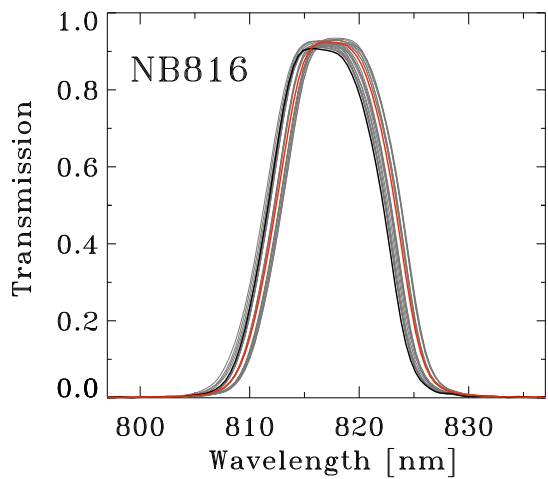} 
\end{center}
\caption{
Filter transmission curves of $NB816$.
The gray lines present the transmission curves at the 21 positions (see text).
The black line is the same as the gray lines, but the transmission curve at the central position.
The red line represents the area-averaged mean transmission
curve that is shown in the Subaru website 
\footnote{
\url{http://www.naoj.org/Observing/Instruments/HSC/sensitivity.html}
}
\label{fig:nb816}}
\end{figure}

\begin{figure}
 \begin{center}
  \includegraphics[width=8cm]{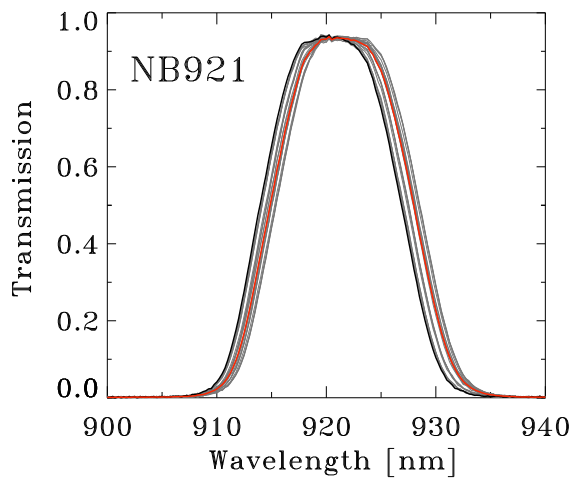} 
\end{center}
\caption{
Same as Figure \ref{fig:nb816}, but for $NB921$.
\label{fig:nb921}}
\end{figure}

Figures \ref{fig:nb816} and \ref{fig:nb921} present the transmission curves of the 
HSC $NB816$ and $NB921$ filters that are used in the HSC first-data 
release \citep{aihara2017}.
We measure the transmission curves at 21 points over the
600mm-diameter circular filter.
The detectors of the camera need transmission of the circular filter
from the center $r=0$ to the radius of $r=280$ mm.
Facing on the filter, we define the angle of 0 deg for the arbitrary direction
from the filter center to the edge. On the angle of 0 deg, we measure
transmission curves at the 5 positions of $r=50$, $100$, $150$, $200$, and $270$ mm.
We change the angle to 90, 180, and 270 deg for the counter clockwise direction,
and obtain the transmission curves at the 20 positions ($=5\times 4$). Including
the measurement at the center ($r=0$ mm), we have the transmission
curves at a total of 21 positions. Note that the measurement position of $NB921$ is 
slightly different from the one of $NB816$. The $r=270$ mm position is replaced with 
$r=265$ mm for the $NB921$ filter. We find that the area-weighted average central wavelength
and FWHM of NB816 (NB921) are 
$817.7$ nm and $11.3$ nm ($921.5$ nm and $13.5$ nm), respectively,
and that the central-position central wavelength and FWHM of NB816 (NB921) are
$816.8$ nm and $11.0$ nm ($920.5$ nm and $13.3$ nm), respectively.
At any positions within $r\le 265$ mm, the peak transmissions of NB816 and NB921 are high, $>90$\%.
The deviations of the central and FWHM wavelengths are typically
within $\simeq 0.3$\% and $\simeq 10$\%, respectively. 
%

Using the HSC narrowband filters, we study LAEs at $z=2.2$, $5.7$ and $6.6$ ($7.3$)
over the large areas of $26$ ($3.5$) deg$^2$ that are
about an order of magnitude larger than those of the previous studies
for a given redshift.
Exploiting samples of $z=2.2$ and $5.7$ LAEs, we study evolution of LAEs 
at $z=2.2-5.7$, from the lowest redshift accessible by ground-based observations ($z\sim 2$)
to the high-$z$ edge of the post reionization epoch ($z\sim 6$). The LAEs at three redshifts of $z=5.7$, $6.6$, and $7.3$
allow us to study evolution of galaxy formation as well as cosmic reionization
with the Ly$\alpha$ damping wing absorption of the neutral IGM.
The LAEs at the post reionization epoch of $z=5.7$ play a role of the baseline
for the properties of LAEs with no cosmic reionization effects given by the IGM Ly$\alpha$ damping wing absorption.
The comparisons of $z=5.7$ and $6.6$ LAEs over the $26$ deg$^2$ sky provide 
large statistical results, while the $z=7.3$ LAEs in the moderately large 3.5 deg$^2$ area 
allows us to investigate galaxy formation and cosmic reionization at the heart of EoR. 
Based on the data sets of the HSC SSP survey (see \cite{aihara2017} for the details of the first-data release)
and our extensive spectroscopic follow up observations with Subaru, Keck, and Magellan telescopes, 
we start the program named
SILVERRUSH (Section \ref{sec:introduction}).
This program is one of the twin programs. 
%
%
The other program
%
%
is the study for dropouts, Great Optically Luminous Dropout Research Using Subaru HSC (GOLDRUSH), 
that is detailed in \citet{ono2017}, \citet{harikane2017}
%
%
, and \citet{toshikawa2017}.
%
%

%
%
In a series of SILVERRUSH papers, we present the SILVERRUSH program strategy (this paper), 
%
%
%
%
%
LAE sample selections \citep{shibuya2017a}, 
and spectroscopic follow-up observations \citep{shibuya2017b}, Ly$\alpha$ LF evolution \citep{konno2017}, 
LAE clustering evolution (this paper), and the comparisons with numerical simulation results \citep{inoue2017}.
These are early papers of the SILVERRUSH program whose observations are still underway.
In these early papers, we present the results of the LAE studies based on the images
taken until 2016 April that include neither $NB387$ (for $z=2.2$ LAEs) nor $NB101$ (for $z=7.3$ LAEs) data, 
but only $NB816$ and $NB921$ data for $z=5.7$ and $6.6$ LAEs, respectively (Section \ref{sec:sample}).
Here, we aim
%
%
that SILVERRUSH results can serve as the baselines of LAE properties 
useful for the on-going LAE studies including MUSE, HETDEX, and the forthcoming $z>7$ galaxy
studies of James Webb Space Telescope (JWST) and extremely large telescopes (ELTs).


\begin{longtable}{cccccccccl}
  \caption{LAE Samples and the Clustering Measurements}\label{tab:clustering_powerlaw}
  \hline              
  $z$ & $L_{\rm Ly\alpha}^{\rm th}$ & $n_{\rm g}$ & $N$ & IC  & $r_0$ & $b_{\rm avg}$ & $r_0^{\rm max}$  & $b_{\rm avg}^{\rm max}$ & $\chi^2/$dof\\    
        &   ($10^{42}$ erg s$^{-1}$)     & ($10^{-5}$ Mpc$^{-3}$)  &        &     ($10^{-3}$)   & ($h_{100}^{-1}$ Mpc) &  &  ($h_{100}^{-1}$ Mpc)  &      &  \\ 
  (1) & (2) & (3) & (4) & (5) & (6) & (7) & (8) & (9) & (10)\\ 
\endfirsthead
  \hline
  Name & Value1 & Value2 & Value3 & Value4 & Value5 & Value6 & Value7 & Value8 & Value9\\
\endhead
  \hline
\endfoot
  \hline
  \multicolumn{10}{l}{\footnotesize For comparison,
$r_0$ is expressed with $h_{100}$. $^\dagger$ The number including sources in D-ELAIS-N1.}\\
\endlastfoot
  \hline
%
          $5.7$ & 6.3 & 7.3 & 959$^\dagger$ & 1.83 & $3.01^{+0.35}_{-0.35}$ & $4.13\pm 0.17$ & $4.47^{+0.52}_{-0.52}$ & $5.90\pm 0.24$ & $5.69/5$\\
          $6.6$ & 7.9 & 2.4 & 873                   & 1.35 & $2.66^{+0.49}_{-0.70}$ & $4.54\pm 0.63$ & $3.95^{+0.73}_{-1.04}$ & $6.49\pm 0.90$ & $2.70/2$\\
\end{longtable}

\section{Observations and Sample}
\label{sec:sample}

%
%
The HSC SSP survey started observations in March 2014,
%
%
taking deep broadband and narrowband images on large areas of the sky (PI: S. Miyazaki).
In our study, we use the HSC SSP survey (HSC et al. in prep) data taken until 2016 April
%
%
with two narrowbands ($NB816$ and $NB921$) and five broadbands ($grizy$)
useful for our LAE studies. 
Because the data of $NB387$ and $NB101$ (for detections of LAEs at $z=2.2$ and $7.3$)
are not taken until 2016 April, we present the results of imaging with the two narrowband
$NB816$ and $NB921$ (for LAEs at $z=5.7$ and $6.6$) whose observations have been 
partly conducted.
%
%
%
%
The data are reduced with the HSC pipeline software,
hscPipe version 4.0.2 \citep{bosch2017}.
%
%
The $5\sigma$ depths of the imaging data are typically
$\simeq 25-25.5$ and $\simeq 26-27$ magnitudes in narrowbands and broadbands, respectively
(see \cite{shibuya2017a} for more details).
Note that the HSC SSP survey data set used in this study 
is notably larger than the one of the first-data release \citep{aihara2017}
that is composed of images taken only in 2014 March and 2015 November.

Our LAEs are selected with the combinations of broadband and narrowband
colors down to 
the $5\sigma$ detection limits.
The LAEs should have a narrowband excess,
the existence of Gunn-Peterson trough, and no detection of blue continuum fluxes.
The color selection criteria are similar to those of \cite{ouchi2008} and \cite{ouchi2010}
that are defined as
\begin{eqnarray}
\label{eq:selectionNB816}
i-NB816\ge1.2\ {\rm and}\ 
g>g_{3\sigma}\ {\rm and}\ \ \ \ \ \ \ \ \ \ \ \ \ \ \ \ \ \ \ \ \ \ \nonumber\\
\ \ \ \ \ \left[(r\le r_{3\sigma}\ {\rm and}\ r-i\ge 1.0)\ {\rm or}\ (r>r_{3\sigma})\right]
\end{eqnarray}
and
\begin{eqnarray}
\label{eq:selectionNB921}
z-NB921\ge1.0\ {\rm and}\ 
g>g_{3\sigma}\ {\rm and}\
r>r_{3\sigma}\ {\rm and}\ \nonumber\\
\ \ \ \ \ \left[(z\le z_{3\sigma}\ {\rm and}\ i-z\ge 1.3)\ {\rm or}\ (z>z_{3\sigma})\right]
\end{eqnarray}
for $z=5.7$ and $6.6$ LAEs, respectively, 
where $g_{3\sigma}$ ($r_{3\sigma}$) is the $3\sigma$ limiting 
magnitude of $g$ ($r$) band that ensures 
no detection of a continuum bluer than the Lyman break.
Similarly, $z_{3\sigma}$ is the $3\sigma$ detection 
limit of $z$ band. After applying the candidate screening on the
basis of hscPipe parameters+flags and visual inspection,
we obtain
a total of 2,354 LAEs
(1,081 and 1,273 LAEs at $z=5.7$ and $6.6$, respectively)
from the LAE {\tt ALL} catalogs \citep{shibuya2017a}.
We investigate the 2,354 LAEs, and make homogeneous samples
over the survey areas that consist of
%
%
%
1,832 LAEs (959 and 873 LAEs at $z=5.7$ and $6.6$, respectively)
with a common narrowband limiting magnitude of 
$NB816<25.0$ or $NB921<25.0$ that is surely brighter than the $5\sigma$ detection levels.
The homogenous samples of $z=5.7$ and $6.6$ LAEs
have very similar threshold luminosities (the rest-frame equivalent width) of 
%
%
$L_{\rm Ly\alpha}^{\rm th}>6.3 \times 10^{42}$ and $7.9 \times 10^{42}$ erg s$^{-1}$ 
($EW_0 \gtrsim 20$\AA) at $z=5.7$ and $6.6$, respectively.
The threshold luminosities correspond to $\simeq L^*$ luminosities 
(e.g. \cite{ouchi2008,ouchi2010,konno2017}).
The LAE sample selection is detailed in \citet{shibuya2017a}.

These LAEs are found in a total of $13.8$ ($21.2$) deg$^2$ area 
consisting of the fields named
D-ELAIS-N1, D-DEEP2-3, UD-COSMOS, and UD-SXDS for the $z=5.7$ LAE sample
(D-ELAIS-N1, D-DEEP2-3, D-COSMOS, UD-COSMOS, and UD-SXDS for the $z=6.6$ LAE sample).
%
%
%
%
Because, for the $z=5.7$ and $6.6$ LAE samples, the redshift ranges are 
$z=5.726\pm 0.046$ and $z=6.580\pm 0.056$
in the case that top-hat selection functions of LAEs with the FWHMs of 
the narrowband filters are assumed,
the total survey volumes covered with the narrowband transmission are
$1.2\times 10^7$ and $1.9\times 10^7$ comoving Mpc$^{3}$ at $z=5.7$ and $6.6$,
respectively.
Note that the total areas of $13.8$ and $21.2$ deg$^2$ for the $z=5.7$ and $6.6$ samples
correspond to 
%
%
0.3 and 0.5 comoving Gpc$^2$ areas, respectively.
These large survey areas allow us to study average properties of LAEs at $z=6-7$ whose observabilities are 
%
%
influenced by patchy ($10-100$ Mpc)
%
%
ionized bubbles
probably existing at the end of the EoR \citep{furlanetto2006}.
%

%
%
%

%
%

%
In our LAE samples, a total of 97 LAEs at $z=5.7$ and $6.6$ are confirmed
by our spectroscopic observations and the cross-matching of the existing spectra \citep{shibuya2017b}. 
Because there are 81 LAE candidates whose spectroscopic identifications are obtained
by our past and present programs, we estimate contamination rates of our LAE samples
with these 81 LAE candidates that consist of 53 and 28 LAEs
at $z=5.7$ and $6.6$, respectively. We find 4 out of 53 (4 out of 28) candidates are
foreground contamination objects,
and estimate the contamination rates to be $\sim 8$\% and $\sim 14$\% for the samples of LAEs at $z=5.7$ and $6.6$,
respectively. We also investigate contamination rates of bright LAEs that are brighter than 
24 magnitudes in the narrowband.
There are 6 and 12 bright LAE candidates with spectroscopic identifications. The spectroscopic results
indicate that 1 out of 6 (4 out of 12) LAE candidates are foreground interlopers,
which correspond to the contamination rate of $\sim 17$\% ($\sim 33$\%) for the sample of LAEs at $z=5.7$ ($6.6$).
Although the contamination rates depend on magnitude, the contamination rates indicated by 
the spectroscopic confirmation range in around $0-30$\% in our $z=5.7$ and $6.6$ LAE samples.
%
%
Although the contamination rates include large uncertainties due to
the small number of our spectroscopically confirmed sources 
at this early stage of our LAE studies, we assume that the contamination
rates are $0-30$\% in our LAE samples that are used in our analysis below.

\begin{figure*}
 \begin{center}
     \includegraphics[width=11cm]{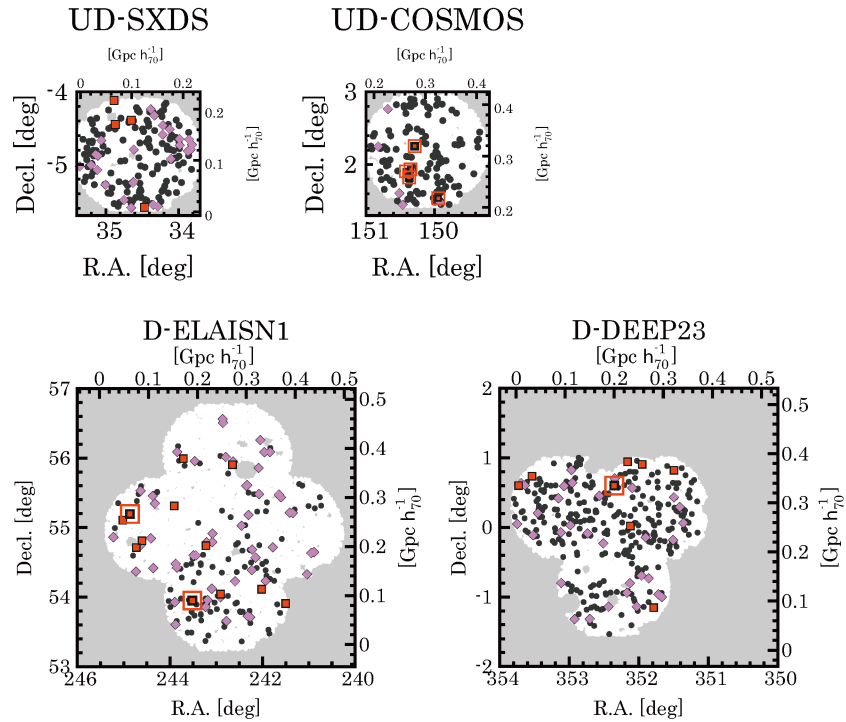} 
\end{center}
\caption{
Sky distribution of the LAEs at $z=5.7$.
The red squares, magenta diamonds, and black circles
represent positions of narrowband bright ($<23.5$ mag), medium-bright ($23.5-24.0$ mag),
and faint ($24.0-25.0$ mag) LAEs, respectively. 
The large red open square indicates the LAEs with spatially extended Ly$\alpha$
emission (\cite{shibuya2017b}).
%
%
%
The gray shades denote either the areas with no HSC data or
the masked regions with a bad data quality.
%
%
The scale on the map is marked in angles (degrees) and 
the projected distances (comoving megaparsecs).
\label{fig:dist_nb816LAE}}
\end{figure*}

\begin{figure*}
 \begin{center}
  \includegraphics[width=17cm]{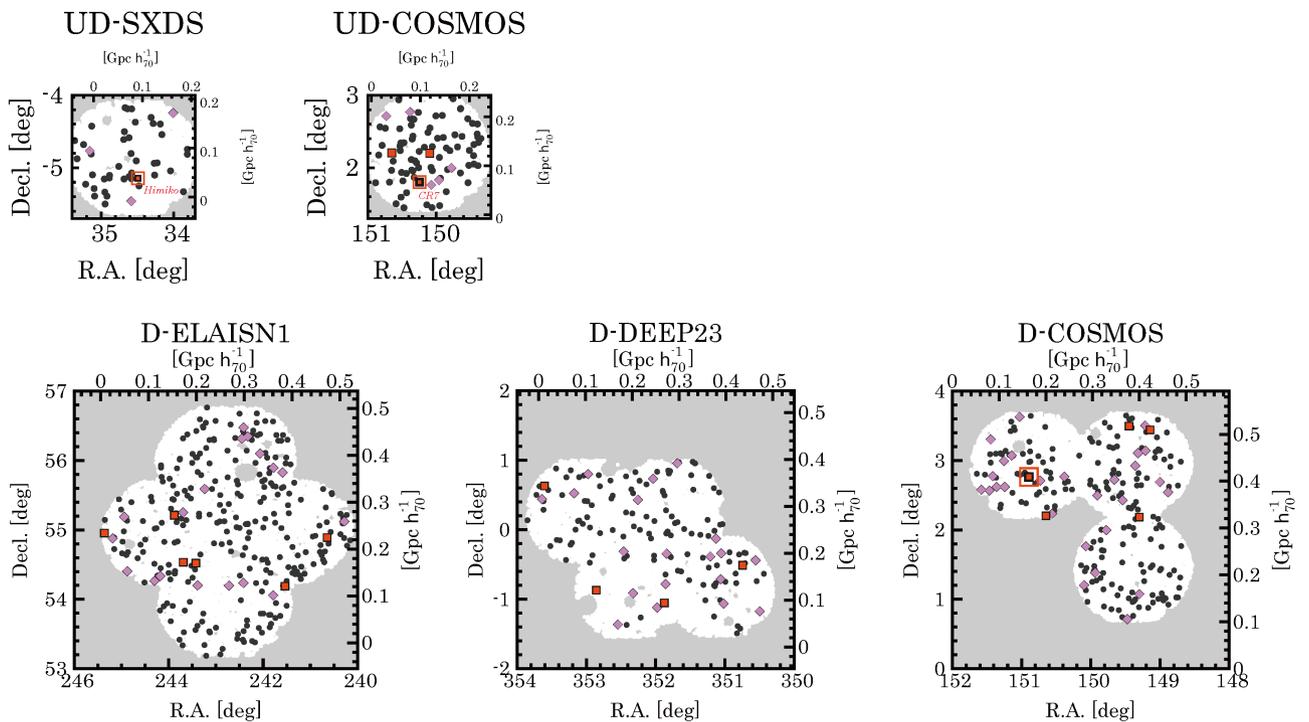} 
\end{center}
\caption{
Same as Figure \ref{fig:dist_nb816LAE}, but for the LAEs $z=6.6$.
The large red open squares indicate the LAEs with spatially extended Ly$\alpha$
emission including Himiko \citep{ouchi2009a} and CR7 \citep{sobral2015}. 
See \citet{shibuya2017b} for more details.
\label{fig:dist_nb921LAE}}
\end{figure*}

\begin{figure*}
 \begin{center}
  \includegraphics[width=16cm]{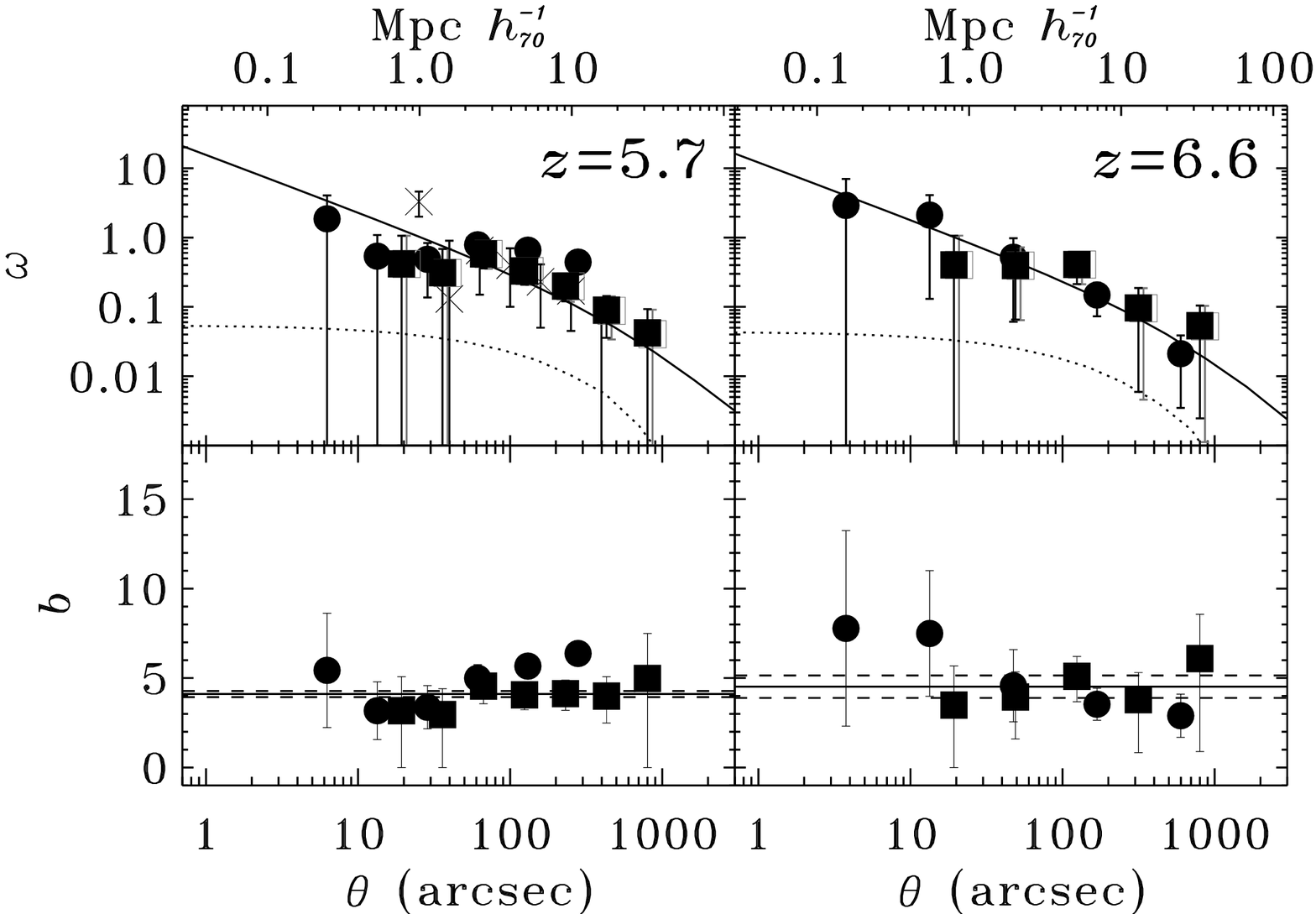} 
\end{center}
\caption{
Angular correlation function (ACF) and bias
of the LAEs at $z=5.7$ (left) and $6.6$ (right).
The top and bottom panels show the ACFs and the bias.
The black-filled and gray-open squares represent the ACFs of
our LAEs with and without the IC correction, respectively,
while the filled circles denote the ACFs with the IC correction derived
by \citet{ouchi2010}.
For presentation purposes,
we slightly shift the gray open squares along the abscissa.
%
In the top panels, the solid and dotted lines 
present the best-fit power-law functions of our ACFs
and the ACFs of the underlying dark matter predicted 
by the linear theory (e.g. \cite{peacock1994}).
Because the power-law spatial correlation function is projected on sky with the method in \citet{simon2007}, the best-fit power-law functions are curves.
In the bottom panels, the solid and dashed lines
indicate the average bias and the $1\sigma$ error values,
respectively. The crosses in the top left panel show
the ACFs obtained by \citet{murayama2007}.
The top axis denotes the projected distance in comoving 
megaparsecs. 
\label{fig:acorr_hscNB816NB921_all}
}
\end{figure*}

\section{Clustering Analysis and Results}
\label{sec:results}

Figures \ref{fig:dist_nb816LAE} and \ref{fig:dist_nb921LAE} present the sky distribution of the LAEs
at $z=5.7$ and $6.6$, respectively. In Figure \ref{fig:dist_nb816LAE},
we find that the density of $z=5.7$ LAEs in the D-ELAIS-N1 field is 
largely biased to the southern region, 
%
%
considering
%
%
the possibilities of the detection incompleteness and the real large overdensity. We have 
carefully investigated our D-ELAIS-N1 images, but found no significant
spatial fluctuations of the detection completeness produced by the inhomogeneity of 
the data handling and the data qualities including depths and seeing sizes.
%
However, there remain the possibilities that unknown effects of the data reduction
may produce the large overdensity. Because the origin of this large 
overdensity is still unknown, we do not use the data of D-ELAIS-N1 in this study.
Instead, we discuss this overdensity in R. Higuchi et al. (in preparation).
%
%
We thus use 
the remaining 734 LAEs at $z=5.7$ in our analysis.

We quantify clustering properties based on the measurements of 
the angular correlation functions in the following sections.



\subsection{Angular Correlation Function}
\label{sec:spatial_angular}

We derive angular two-point correlation functions (ACFs)
of our LAEs in the same manner as 
\citet{ouchi2003}, \citet{ouchi2005b}, \citet{ouchi2010}, and \citet{harikane2016}.
We use the estimator of \citet{landy1993},
\begin{equation}
\omega_{\rm obs}(\theta)
  = \frac{DD(\theta)-2DR(\theta)+RR(\theta)}{RR(\theta)},
\label{eq:landyszalay}
\end{equation}
where $DD(\theta)$, $DR(\theta)$, and $RR(\theta)$ are numbers of
galaxy-galaxy, galaxy-random, and random-random pairs normalized by
the total number of pairs in each of the samples of the pairs.
%
%
We use the random catalog (Coupon et al. in preparation) whose 
surface number density is $100\ \mathrm{arcmin}^{-2}$.
The random catalog has the geometrical constraint same as 
the one of our LAEs, representing our survey areas.
%
%
Statistical errors are estimated 
%
with Jackknife 
%
%
resampling with subsamples each of which has a $\sim1000^2\ \mathrm{arcsec}^2$ area.
Figure \ref{fig:acorr_hscNB816NB921_all} shows the observed ACFs 
$\omega_{\rm obs}(\theta)$ of our LAEs at $z=5.7$ and $6.6$.
The ACF measurements cover the scale of $\sim 1-100$ comoving Mpc
that is indicated in the upper axes of Figure \ref{fig:acorr_hscNB816NB921_all}.
We find that these ACFs are consistent with those obtained with the previous
Subaru/Suprime-Cam data \citep{ouchi2010}, and that 
our present HSC data provide the large-scale ACFs with uncertainties
smaller than those of the previous data.

Although our survey area is large, we evaluate 
%
%
%
the integral constraint IC \citep{groth1977}.
The IC value corresponds to the observational offset
in $\omega_{\rm obs}(\theta)$ originated by 
a limited survey area.
%
%
Including the correction for the number of objects in the sample, $N$, the true ACF is given by
\begin{equation}
\omega(\theta)=\omega_\mathrm{obs}(\theta)+\mathrm{IC}+\frac{1}{N},
\end{equation}
We evaluate the integral constraint with 
\begin{equation}
\mathrm{IC}=\frac{\Sigma_i RR(\theta_i)\omega(\theta_i)}{\Sigma_i RR(\theta_i)},
\end{equation}
We adopt the model ACF of \citet{simon2007} for the function of $\omega(\theta)$
that is detailed in Section \ref{sec:spatial_correlation_bias}.
%
%
%
We show IC values in Table \ref{tab:clustering_powerlaw}, and
plot the ACFs corrected for IC
in Figure \ref{fig:acorr_hscNB816NB921_all}.
%
%
Note that the IC values are very small, because our survey areas
are very large.
%
%

A clustering signal of galaxies is diluted by contamination
objects in a galaxy sample. If the galaxy sample includes
randomly-distributed contamination objects with a fraction $f_{\rm c}$,
the value of the ACF is reduced by a factor of $(1 - f_{\rm c})^2$.
The ACF corrected for
the randomly-distributed contamination ${\omega}_{\rm max}$
is written as
\begin{equation}
\omega_{\rm max} = \frac{\omega}{(1-f_{\rm c})^2}.
\label{eq:clustering_dilution}
\end{equation}
This is the maximum reduction of the ACF, because 
the real contamination objects are not randomly
distributed but spatially correlated. 
For reference, we evaluate the possible maximum values of $\omega_\mathrm{max}$,
using the contamination fraction whose upper limit is 
$f_{\rm c}=0.3$ (Section \ref{sec:sample}).







\subsection{Correlation Length and Bias}
\label{sec:spatial_correlation_bias}

We fit the ACFs with a simple power law model of the spatial correlation function,
\begin{equation}
\xi(r)=\left(\frac{r}{r_0}\right)^{-\gamma},
\label{eq:3dcorrelation}
\end{equation}
where $\gamma$, $r_0$, and $r$ are 
the slope of the power law, the correlation length,
and the spatial separation between two galaxies,
respectively.
The spatial correlation function is related to the ACF 
with the Limber equation
\citep{peebles1980,efstathiou1991}
%
%
that is an integral equation of the (three-dimensional) spatial correlation
function connecting with the (two-dimensional) ACF.
%
%
However, \citet{simon2007} claims that the Limber equation
does not provide accurate values in a very large separation of galaxies
whose redshift-distribution distance is narrower than the transverse distance
in the case such for narrowband-selected LAEs.
We thus adopt the method that \citet{simon2007} suggests,
and derive $r_0$ of our LAEs, fitting the power-law functions
to our ACFs. 
Because no meaningful constraints on $\gamma$ are obtained with
the present samples, we adopt the fiducial $\gamma$ value of $\gamma=1.8$
%
%
that is adopted in the previous clustering analyses for LAEs
\citep{ouchi2003,gawiser2007,kovac2007,ouchi2010}.
To investigate the dependences of our results on the value of $\gamma$,
we also use the other fixed $\gamma$ values of $\gamma=1.6$ and $2.0$
that bracket the possible range of the power-law index of high-$z$ galaxies 
at $z\sim 4-6$ \citep{lee2006,mclure2009}.
We have found that neither $r_0$ nor bias values changes over 10\%, 
and that $r_0$ and bias values fall well within the errors.
%
%
Here, we use the redshift distribution of our LAEs 
estimated with the narrowband transmission curve.
The best-fit power-law functions are presented
in Figure \ref{fig:acorr_hscNB816NB921_all}.
The correlation lengths thus obtained are 
$r_0=3.01^{+0.35}_{-0.35}$ and $2.66^{+0.49}_{-0.70}$ $h^{-1}_{100}$ Mpc
for LAEs at $z=5.7$ and $6.6$, respectively, that are
summarized in Table \ref{tab:clustering_powerlaw}.
%
To make comparisons with the previous results, we express
the correlation lengths with $h_{100}$, the Hubble constant in
units of 100 km s$^{-1}$, instead of 70 km s$^{-1}$.
Table \ref{tab:clustering_powerlaw} also shows
the upper limit of the correlation length, $r_0^{\rm max}$, 
that is given by
\begin{equation}
r_0^{\rm max}=r_0\left(\frac{1}{1-f_\mathrm{C}}\right)^{2/\gamma}.
\end{equation}
We obtain $r_0^{\rm max} = 4.47^{+0.52}_{-0.52}$ and $3.95^{+0.73}_{-1.04}$ $h_{100}^{-1}$ Mpc 
for our $z=5.7$ and $6.6$ LAEs, respectively.

These correlation length values are consistent with the previous measurements within the errors.
\citet{ouchi2010} obtain ($r_0$, $r_0^{\rm max}$) of 
($3.12^{+0.33}_{-0.36}$, $4.30^{+0.45}_{-0.50}$) and ($2.31^{+0.65}_{-0.85}$, $3.60^{+1.02}_{-1.32}$) 
for LAEs at $z=5.7$ and $6.6$, respectively.
%

\begin{figure}
 \begin{center}
  \includegraphics[width=8cm]{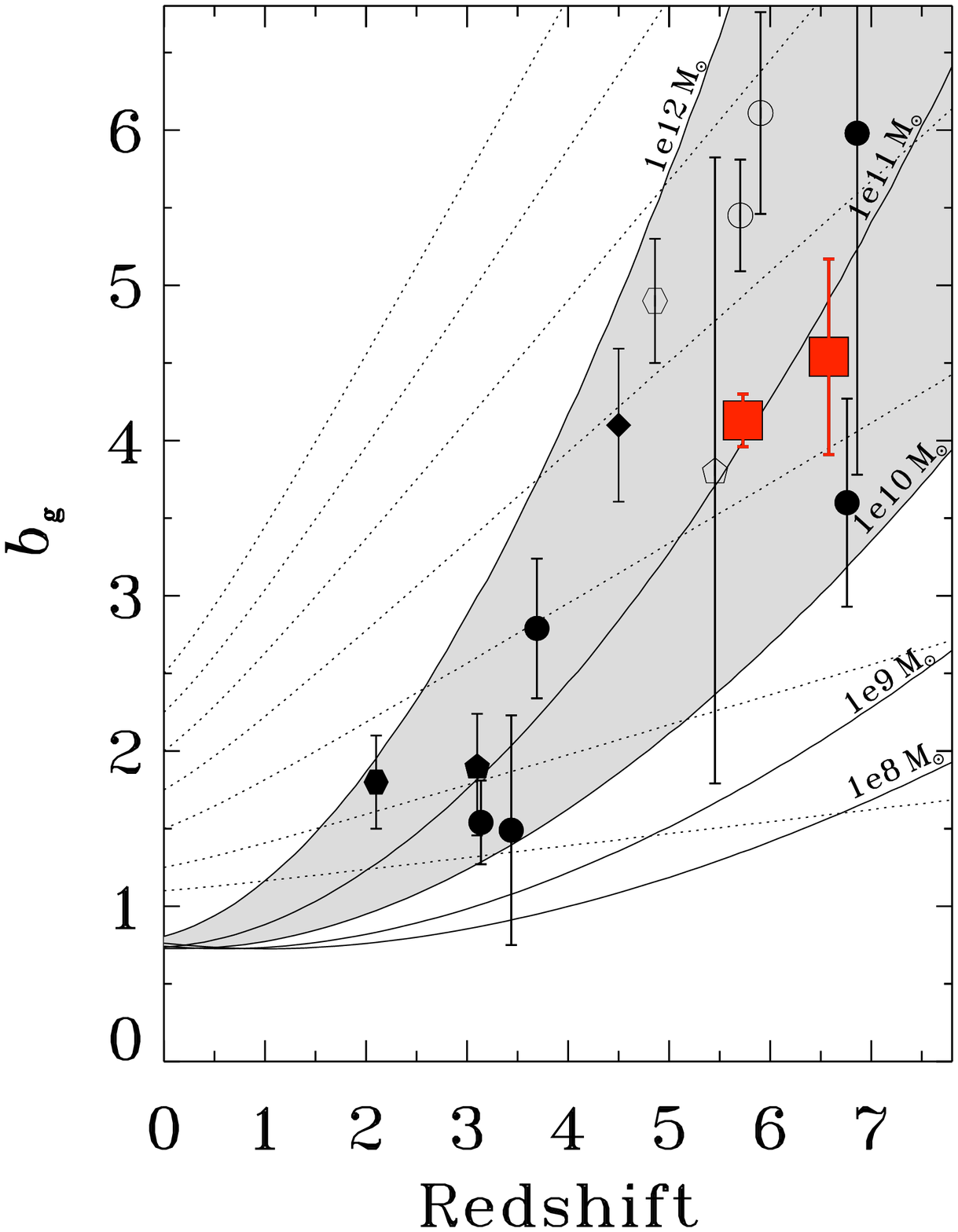}
\end{center}
\caption{
Bias of $\gtrsim L^*$ LAEs as a function of redshift.
The red squares represent bias of our LAEs at $z=5.7$ and $6.6$,
while the filled circles denote those of LAEs obtained in 
\citet{ouchi2010}. The open circles, pentagon, and hexagon
indicate LAE bias values obtained by \citet{ouchi2010}, \citet{ouchi2005a},
and \citet{ouchi2003}, respectively.
The filled hexagon, pentagon, and diamond are LAE bias
values measured by \citet{guaita2010}, \citet{gawiser2007},
and \citet{kovac2007}, respectively.
%
For the previous study results, we correct for the difference of $\sigma_8$.
To clarify the bias measurements, we give small offsets along 
the abscissa axis to the data points of the previous studies.
The solid lines indicate bias of dark-matter halos
with a halo mass of $10^{8}$, $10^{9}$, $10^{10}$, $10^{11}$,
and $10^{12} M_\odot$ in the case of one-to-one 
correspondence between galaxies and dark-matter halos \citep{ouchi2010}.
The gray region shows the dark-matter halo mass range of $10^{10}-10^{12} M_\odot$.
The dotted lines are evolutionary tracks of bias 
in the case of the galaxy-conserving model (eq. \ref{eq:galaxy_conserving}).
\label{fig:redshift_bias_LBGLAE}
}
\end{figure}

In the framework of the $\Lambda$CDM model,
the galaxy-dark matter bias is defined by
\begin{equation}
b_{\rm g}(\theta)^2
 \equiv
  \omega(\theta)/\omega_{\rm dm}(\theta),
\label{eq:bias}
\end{equation}
where $\omega_{\rm dm}(\theta)$ is the 
ACF of dark-matter predicted 
by the linear theory model (e.g. \cite{peacock1994}).
The model calculations are 
made with the same survey volumes as 
those of our LAE samples.
Note that $b_{\rm g}$ is the bias value equivalent
to the one given by the three-dimensional
spatial correlation functions, 
$b_{\rm g}^2=\xi(r)/\xi_{\rm dm}(r)$,
where $\xi_{\rm dm}(r)$ is the spatial correlation function
of dark matter.

In Figure \ref{fig:acorr_hscNB816NB921_all},
the top and bottom panels show $\omega_{\rm dm}(\theta)$
and $b_{\rm g}(\theta)$, respectively.
We estimate the average bias $b_{\rm avg}$,
averaging $b_{\rm g}(\theta)$ with error weighting 
over the angular range presented in Figure \ref{fig:acorr_hscNB816NB921_all}.
We also calculate the upper limits of bias values $b_{\rm avg}^{\rm max}$
with the maximal contamination correction that corresponds 
to $A_{\omega}^{\rm max}$ (eq. \ref{eq:clustering_dilution}; see Section \ref{sec:spatial_angular}).
We obtain
$b_{\rm avg} =4.13\pm 0.17$ and $4.54\pm 0.63$
( $b_{\rm avg}^{\rm max} = 5.90\pm 0.24$ and $6.49\pm 0.90$)
%
%
%
for our $z=5.7$ and $6.6$ LAEs, respectively.
We summarize $b_{\rm avg}$ and $b_{\rm avg}^{\rm max}$ values thus obtained
in Table \ref{tab:clustering_powerlaw}.
We compare these bias values with those obtained in the previous studies.
We find that the bias value of our LAEs are consistent with those 
derived by \citet{ouchi2005a} and \citet{ouchi2010} within the $\sim 1\sigma$ errors.

Figure \ref{fig:redshift_bias_LBGLAE} presents
bias of LAEs as a function of redshift.
In Figure \ref{fig:redshift_bias_LBGLAE},
we plot bias measurements of LAEs at $z=2-7$ 
derived in the previous studies.
We have corrected these previous study measurements of $b_{\rm g}$
for the difference of $\sigma_8$ \citep{ouchi2010},
and shown the corrected $b_{\rm g}$ values in Figure \ref{fig:redshift_bias_LBGLAE}.

%
%
This work and the previous studies measure bias
of LAEs whose Ly$\alpha$-luminosity detection limits are similar,
a few times $10^{42}$ erg s$^{-1}$
corresponding to $\sim L^*$ luminosities
%
%
over $z\sim 2-7$ (e.g. \cite{ouchi2008,ouchi2010,konno2016,konno2017}).
%
%
%
%
Thus, we can omit the luminosity segregation effects 
in bias that depends on Ly$\alpha$ luminosity. 
%
%
Figure \ref{fig:redshift_bias_LBGLAE} indicates that the bias of $\gtrsim L^*$ LAEs
significantly increases from $z\sim 2-3$ to $z\sim 6-7$. The physical origins of this increase
are discussed in Section \ref{sec:discussion}.

\begin{figure}
 \begin{center}
 \includegraphics[width=8cm]{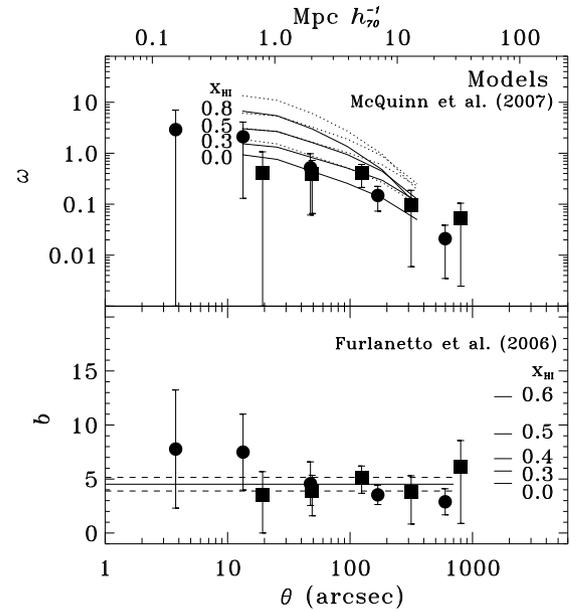} 
\end{center}
\caption{
%
%
%
Comparisons of the theoretical models with the ACF and bias measurements at $z=6.6$.
In the top (bottom) panel,
the filled squares and circles denote the ACF (bias) measurements with the IC correction 
obtained by this work and \citet{ouchi2010}, respectively, 
which are the same as those shown in the top (bottom) right panel of Figure \ref{fig:acorr_hscNB816NB921_all}.
$Top:$ The solid and dotted lines represent the \authorcite{mcquinn2007}'s (\yearcite{mcquinn2007}) models of 
$z=6.6$ LAEs with a dark-matter halo masses of $3\times 10^9 M_\odot$ and $3\times 10^{10}M_\odot$, respectively.
The model of $3\times 10^9 M_\odot$ is the modification of the model of $3\times 10^{10}M_\odot$ (see text).
From the bottom to the top solid (dotted) lines, neutral hydrogen fractions of the IGM are
$x_{\rm HI}=0.0$, $0.3$, $0.5$, and $0.8$.
%
%
$Bottom:$ The ticks at the right-hand side
indicate bias values predicted by \authorcite{furlanetto2006}'s (\yearcite{furlanetto2006}) 
models 
%
%
(see text). The solid and dashed lines indicate the average bias and the $1\sigma$ uncertainty values, respectively,
that are the same as those shown in the bottom right panel of Figure \ref{fig:acorr_hscNB816NB921_all}.
%
%
\label{fig:acorr_NB921_model}
}
\end{figure}

\section{Discussion}
\label{sec:discussion}

\subsection{Cosmic Reionization}
\label{sec:cosmic_reionization}

\subsubsection{Constraints on $x_{\rm HI}$ from the Clustering Results}
\label{sec:constraints_clustering}


Clustering signals of observed LAEs are strengthened
%
%
by the patchy distribution of the neutral hydrogen at the EoR,
because
%
%
Ly$\alpha$ photons of LAEs in the ionized bubbles
can selectively escape from the patchy neutral IGM
with a small amount of Ly$\alpha$ damping wing absorption
(\cite{furlanetto2006,mcquinn2007,lidz2009,iliev2008,sobacchi2015}).
Based on the fact that the Gunn-Peterson optical depth largely increases at $z\sim 6$
\citep{fan2006}, one would expect that the clustering amplitude of the observed LAEs 
increases from $z=5.7$ (the ionized universe) to $z=6.6$ (the partly-neutral universe).
Figure \ref{fig:redshift_bias_LBGLAE} indicates 
no significant rise of bias from $z=5.7$ to $6.6$
beyond the moderately small errors (i.e. by a factor of $\sim 20$\%).
This result suggests that clustering of $z=6.6$ LAEs is not
largely affected by the cosmic reionization effects,
where the bias evolution of the hosting dark-matter halos towards high-$z$ may be 
also involved. Based on this bias evolution result,
we place constraints on cosmic reionization parameters
with the help of theoretical models. We compare our 
observational results with multiple theoretical models,
%
%
because there is a possibility that our conclusions may
depend on a model chosen for the comparisons.
We thus try to avoid model-dependent conclusions as much as possible,
and to obtain more objective interpretations
for our observational results. 
%
%
Note that the arguments below
follow those of \citet{ouchi2010} with our HSC clustering measurements.

In the top panel of Figure \ref{fig:acorr_NB921_model},
we compare our $z=6.6$ LAE clustering measurements 
with those of theoretical predictions \citep{mcquinn2007,furlanetto2006}.
The model of \citet{mcquinn2007} is presented in
the top panel of Figure \ref{fig:acorr_NB921_model}.
\citet{mcquinn2007} conduct radiative transfer simulations predicting
clustering of LAEs at $z=6.6$. Their models assign a Ly$\alpha$ flux
to a dark-matter halo whose mass is beyond a minimum dark-matter halo mass.
%
%
%
We use the minimum dark-matter halo mass of our LAEs at the post-reionization epoch of $z=5.7$
to estimate the intrinsic LAE clustering at $z=6.6$, assuming no redshift evolution ($z=5.7-6.6$) of the minimum dark-matter halo
mass (Section \ref{sec:sample}).
Although the minimum dark-matter halo mass is $3\times 10^{9} M_\odot$ for the $z=5.7$ LAEs (Section \ref{sec:halo_mass}),
\citet{mcquinn2007} do not calculate the models of the minimum dark-matter halo mass
as low as $3\times 10^{9} M_\odot$. We thus correct the angular correlation functions of 
the McQuinn et al.'s lowest mass ($3\times 10^{10} M_\odot$) model 
for the difference between $3\times 10^{10}$ and $3\times 10^9 M_\odot$,
using eq. (\ref{eq:bias}).
Here we calculate the bias values of the $3\times 10^{10}$ and $3\times 10^9 M_\odot$ cases
with the best-fit halo model (Section \ref{sec:halo_mass}) 
whose parameters are exactly the same as those determined at $z=5.7$. 
In the top panel of Figure 8, we show the McQuinn et al.’s models for $3\times 10^9 M_\odot$ 
with the correction, together with the original $3\times 10^{10} M_\odot$ model.
Because the McQuinn et al.’s models do not correct for the integral constraint,
we do not compare the model predictions below $\omega \sim 0.1$, where
the contribution of the integral-constraint correction term is large.
The top panel of Figure 8 indicates that our $z=6.6$ LAE data points 
fall in the range of $x_{\rm HI}\simeq 0-0.3$, indicating $x_{\rm HI} =0.15 \pm 0.15$. 
%
%
%

\begin{figure}
 \begin{center}
  \includegraphics[width=8cm]{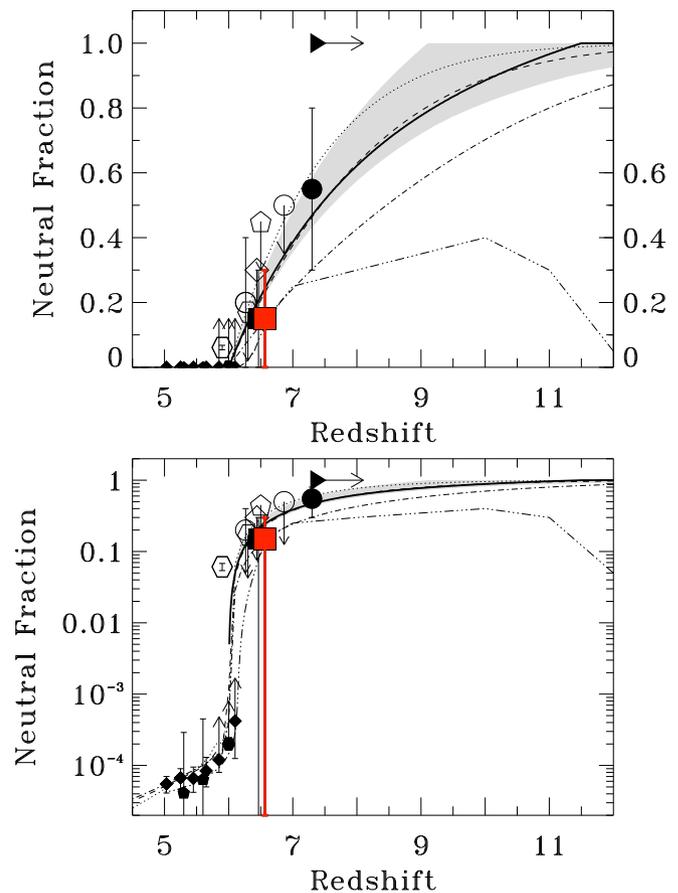} 
\end{center}
\caption{
Neutral hydrogen fraction $x_{\rm HI}$
of the IGM as a function of redshift.
Note that the top and bottom panels are the same,
but with an ordinate axis of linear and log scales, respectively.
The red filled square is the $x_{\rm HI}$ estimate
obtained by our HSC LAE clustering analysis.
The black filled square and circle are the $x_{\rm HI}$ estimates
from the LAE LF evolution of \citet{konno2017} and \citet{konno2014},
respectively.
The open circles are the constraints at $z=6.6$ obtained 
by \citet{ouchi2010} from the evolution of 
Ly$\alpha$ LF (left circle) and clustering (right circle), while
the open diamond and the open pentagon
represent the upper limits from the Ly$\alpha$ LF evolution to $z=6.5$
given by \citet{malhotra2004} and \citet{kashikawa2006}.
Here, we add small offsets along redshift
to the positions of the filled square, the open circles, and
the open diamond, avoiding overlapping symbols.
The filled hexagon and the filled pentagons show
the constraints from a spectrum of a GRB \citep{gallerani2008b}
and statistics of QSO dark-gaps \citep{gallerani2008a},
respectively.
The open hexagons are the constraints calculated
from the Ly$\alpha$ damping wing absorption of 
GRBs at $z=6.3$ \citep{totani2006} and $z=5.9$ \citep{totani2016}.
The filled diamonds indicate the QSO Gunn-Peterson optical depth measurement
results \citep{fan2006}. The triangle denotes
the $1\sigma$ lower-limit of redshift 
obtained by Planck 2015 \citep{planck2016}
in the case of instantaneous reionization.
The solid line and the gray shade indicate
the best-estimate and the uncertainty of 
the $x_{\rm HI}$ evolution \citep{ishigaki2017}
that agrees the evolutions of $\tau_{\rm e}$ and
$\rho_{\rm UV}$ with free parameters including
the ionizing photon escape fraction.
The dotted, dashed, and dot-dashed lines are the
evolution of $x_{\rm HI}$ for the reionizing sources
down to the massive halos, 
the moderately massive halos, and 
the mini-halos, respectively, in the model of \citet{choudhury2008}. 
The dashed double-dotted line indicates the prediction of the
double reionization model \citep{cen2003}.
\label{fig:z_xHI_HSC}
}
\end{figure}

In the bottom panel of Figure \ref{fig:acorr_NB921_model},
we compare our bias results with those of analytical models
of \citet{furlanetto2006} at $z=6.6$. In this comparison, we adopt the bias values
of the small-scale ($\simeq 1-10$ Mpc) of \citet{furlanetto2006}
that is similar to the major angular scale of our bias measurements.
%
%
%
%
%
For the intrinsic bias value of the $z=6.6$ LAEs, 
we use $b=4.6$ 
that is obtained under the assumption of
no evolution from $z=5.7$ to $6.6$ 
for the minimum dark-matter halo mass ($3\times 10^9 M_\odot$)
and the halo-model parameters (Section \ref{sec:halo_mass}).
The bottom panel of Figure \ref{fig:acorr_NB921_model}
presents that our average-bias measurement agrees with the low $x_{\rm HI}$ 
models of \citet{furlanetto2006},
and suggests $x_{\rm HI}\lesssim 0.13$ including the uncertainty of
the average-bias measurement of the $z=6.6$ LAEs.
%
%

%

%
%
The comparisons of our observational results with the models of \citet{mcquinn2007} and \citet{furlanetto2006}
indicate that the neutral hydrogen fraction at $z=6.6$ falls
in the range of $x_{\rm HI}=0.15^{+0.15}_{-0.15}$.
%
%

\begin{figure*}
 \begin{center}
  \includegraphics[width=16cm]{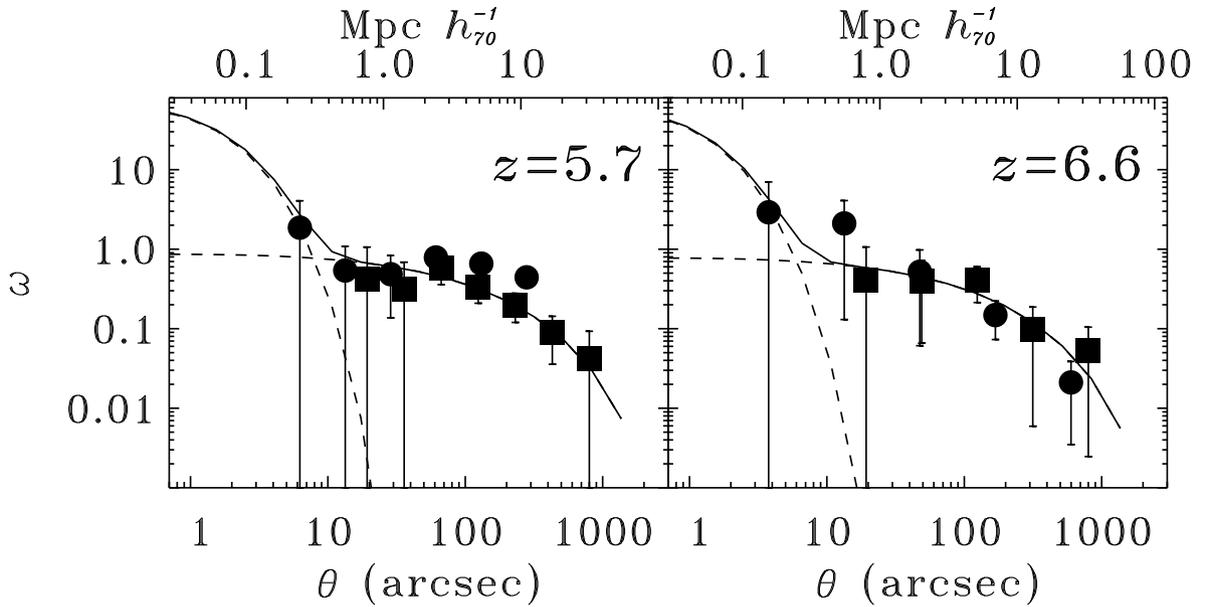} 
\end{center}
\caption{
Best-fit HOD models of the LAEs at $z=5.7$ (left) and $6.6$ (right).
The solid lines represent the best-fit HOD models, while
the dashed lines denote the breakdowns of the best-fit HOD models,
1 and 2 halo terms. The squares and circles are the ACF measurements 
that are the same as those in Figure \ref{fig:acorr_hscNB816NB921_all}.
\label{fig:acorr_hscNB816NB921_HOD}
}
\end{figure*}

\subsubsection{Cosmic Reionization History}
\label{sec:cosmic_reionization_history}

Figure \ref{fig:z_xHI_HSC} shows our $x_{\rm HI}$ value of 
%
%
%
$x_{\rm HI}=0.15^{+0.15}_{-0.15}$
%
%
that is estimated with our LAE clustering measurements. 
This $x_{\rm HI}$ estimate suggests a moderate neutral hydrogen
fraction at $z=6.6$.
In Figure \ref{fig:z_xHI_HSC}, we also present various results of $x_{\rm HI}$
constrained by the similar LAE clustering analysis \citep{ouchi2010}
and the independent $x_{\rm HI}$ estimates from
QSO Gunn-Peterson absorptions \citep{fan2006}
as well as Ly$\alpha$ damping wing (DW) absorption measurements of
QSOs \citep{mortlock2011,bolton2011}, Gamma-ray bursts (GRBs; \cite{totani2006,totani2014,totani2016}; see also \cite{greiner2009}),
and galaxies including LAEs \citep{malhotra2004,ouchi2010,ota2010,kashikawa2011,pentericci2011,
ono2012,treu2013,caruana2014,konno2014,schenker2014,konno2017}.
%
%
%
%
%
Our results agree with the previous
LAE clustering results of \citet{ouchi2010}.
%
%
%
Our constraint on $x_{\rm HI}$ is
stronger than the one of \citet{ouchi2010},
due to the fact that the statistical uncertainties of the ACF measurements of this study
are smaller than those of \citet{ouchi2010}.
This is because the number of LAEs in this study
is larger than the one of \citet{ouchi2010}.
%
%
Moreover, our $x_{\rm HI}$ estimate agrees with
many of the estimates and constraints obtained by the
other independent objects (QSOs, GRBs, and galaxies) and 
methods (Gunn-Peterson and Ly$\alpha$ DW absorptions) 
in the previous studies.
It should be noted that the results of GRBs have a large
scatter. This scatter is probably made by the systematic
uncertainty raised by the sightline variance effects. 
The numerical simulations of \citet{mcquinn2008} suggest 
that the patchiness of reionization gives the systematic error
of $\delta x_{\rm HI}\sim 0.3$ in 
an $x_{\rm HI}$ measurement of a single GRB sightline.
%
%
%
%

\begin{longtable}{ccccccc}
  \caption{Properties of Dark-Matter Halos of the LAEs Estimated with the HOD Model}\label{tab:clustering_hod}
  \hline              
  $z$ & $L_{\rm Ly\alpha}^{\rm th}$ & $\log(M_{\rm min}/M_\odot$)  &  $\log(\left < M \right>/M_\odot)$ & $b_{\rm g}^{\rm HOD}$ & $f_{\rm duty}^{\rm LAE}$ & $\chi^2/$dof\\    
        & ($10^{42}$ erg s$^{-1}$)   &                  &                 &                                 &       ($10^{-4}$)              &       \\ 
  (1) & (2) & (3) & (4) & (5) & (6) & (7)\\ 
\endfirsthead
  \hline
  Name & Value1 & Value2 & Value3 & Value4 & Value5 & Value6 \\
\endhead
  \hline
\endfoot
  \hline
   \multicolumn{7}{l}{\footnotesize 
   $^\dagger$ The results of $z=6.6$ LAEs may be weakly affected by the reionization effects (see text).}\\
\endlastfoot
  \hline
    $5.7$ & $6.3$ & $9.5^{+0.5}_{-1.2}$ & $11.1^{+0.2}_{-0.4}$ & $3.9^{+0.7}_{-1.0}$ &  $7.7^{+53.0}_{-7.6}$ & $1.1/6$\\
    $6.6^\dagger$ & $7.9$ & $9.1^{+0.7}_{-1.9}$ & $10.8^{+0.3}_{-0.5}$ & $4.1^{+1.0}_{-1.4}$ & $0.9^{+13.5}_{-0.9}$ & $1.1/4$\\
\end{longtable}

\subsection{Hosting Dark-Matter Halo}
\label{sec:halo_mass}

Figure \ref{fig:z_xHI_HSC} presents evolution of $x_{\rm HI}$
determined by \citet{ishigaki2017} who use all of the
observational results related to reionization,
CMB Thomson scattering optical depth $\tau_{\rm e}$, 
UV luminosity density $\rho_{\rm UV}$, and
the ionized fraction $Q_{\rm HII}$ obtained to date,
based on the standard analytical model
\citep{madau1999,robertson2015} fitting that
allows a wide parameter range of unknown parameters
such as the ionizing-photon escape fraction.
It should be noted that the \authorcite{ishigaki2017}'s (\yearcite{ishigaki2017}) result of 
$x_{\rm HI}$ evolution 
agrees with the $\tau_{\rm e}$ measurement of Planck2016.
We find in Figure \ref{fig:z_xHI_HSC} that the evolution of $x_{\rm HI}$ (the solid curve)
is consistent with our $x_{\rm HI}$ estimate from the LAE clustering,
supportive of the moderately late reionization scenario suggested by \citet{ishigaki2017}.

\subsubsection{Halo Occupation Distribution Modeling}
\label{sec:HOD_model}
We carry out halo occupation distribution (HOD) modeling for our ACFs
that are the encodes of the hosting dark-matter halo properties.
Note that the ACF of the LAEs at $z=5.7$ is not affected by the effects of
patchy ionized bubbles of cosmic reionization (Section \ref{sec:introduction}), 
because $z=5.7$ corresponds to the post-reionization epoch.
However, the LAE ACF at $z=6.6$ may be strengthened by  
the selective escapes of Ly$\alpha$ photons from patchy ionized bubbles probably
existing at this redshift, although the $z=6.6$ LAE ACF appears weakly 
to be affected by cosmic reionization (Section \ref{sec:constraints_clustering}).
In this section, we perform the HOD modeling for the ACF of the LAEs at $z=5.7$
for the secure results. Moreover, we conduct the same analysis for the LAEs at $z=6.6$ 
under the assumption that the reionization effect is negligibly small, 
following our results that the influence of cosmic reionization is not large at $z=6.6$
(Section \ref{sec:constraints_clustering}).

We adopt the HOD models of \citet{harikane2016} based on the
halo mass function of \citet{behroozi2013} that is a modification of the 
\citet{tinker2008} halo mass function. In the HOD models of \citet{harikane2016},
there are a total of five parameters,
$M_{\rm min}$, $M_1'$,  $\alpha$, $M_0$, and $\sigma_{\rm \log M}$.
Here, $M_{\rm min}$ is the mass of the dark-matter halo that hosts a galaxy 
at the possibility of 50\%. The values of $M_1'$ and $\alpha$ are the normalization and the slope
of the power law, respectively, for the satellite galaxy occupation number.
We define $M_0$ and $\sigma_{\rm \log M}$ as 
the halo mass cut off and the halo-mass transition width of the central and satellite galaxies, respectively.
We do not include the parameter of the duty cycle in our HOD modeling, because
the number density is not used in our fitting. Instead, we compare 
the number densities given by our HOD modeling and our observations
to constrain the duty cycle of LAEs, which are detailed in Section \ref{sec:duty_cycle}.
Following \citet{harikane2016}, we adopt the two relations,
$\log M_0 = 0.76 \log M_1' + 2.3$ and $\log M_1'=1.18 \log M_{\rm min} - 1.28$.
We assume $\sigma_{\rm \log M}=0.2$ and $\alpha=1$ that are 
suggested by the fitting results of \citet{zheng2007}.
although we confirm that the choice of these two parameters do not change our conclusions.
We thus obtain the best-fit HOD models (i.e. $M_{\rm min}$ values) by 
$\chi^2$ minimization. Figure \ref{fig:acorr_hscNB816NB921_HOD} presents
the best-fit HOD models, and Table \ref{tab:clustering_hod} summarizes the
best-fit $M_{\rm min}$ values. We define the average dark-matter halo mass 
$\left < M_{\rm h} \right >$ and the average galaxy bias $b_{\rm g}^{\rm HOD}$ from
the HOD modeling results,
\begin{eqnarray}
\left < M_{\rm h} \right > & = & \frac{\int_{M_{\rm h}^{\rm min}}^{\infty}\ M N_{\rm g}\ n\ dM}{\int_{M_{\rm h}^{\rm min}}^{\infty}\ N_{\rm g}\ n\ dM}\ \ \ {\rm and}
\label{eq:average_halo_mass} \\
b_{\rm g}^{\rm HOD} & = & \frac{\int_{M_{\rm h}^{\rm min}}^{\infty}\ b\ N_{\rm g}\ n\ dM}{\int_{M_{\rm h}^{\rm min}}^{\infty}\ N_{\rm g}\ n\ dM},
\label{eq:average_bias}
\end{eqnarray}
where $n$ and $b$ are the number density and the bias of dark-matter halos, respectively,
with a dark-matter halo mass $M_h$ \citep{behroozi2013,tinker2010}. The value of $N_{\rm g}$ is 
a galaxy occupation function at $M_h$
that is determined by the HOD modeling.
The values of $\left < M_{\rm h} \right >$ and $b_{\rm g}^{\rm HOD}$ for our LAEs 
are summarized in Table \ref{tab:clustering_hod}.
We find that the average dark-matter halo masses of the $\gtrsim L^*$ LAEs 
are moderately small,
$\log (\left < M_{\rm h} \right > /M_\odot) = 11.1^{+0.2}_{-0.4}$ at $z=5.7$
and
$\log (\left < M_{\rm h} \right > /M_\odot) = 10.8^{+0.3}_{-0.5}$ at $z=6.6$
that are consistent with the previous estimates ($\log (\left < M_{\rm h} \right > /M_\odot) \sim 10-11$) 
by clustering analysis \citep{ouchi2010}.
The values of $b_{\rm g}^{\rm HOD}$ are consistent with those of $b_{\rm avg}$ (Table \ref{tab:clustering_powerlaw}) 
within the errors.

\subsubsection{Duty Cycle}
\label{sec:duty_cycle}


It is suggested that LAEs do not exist in all of the dark-matter halos more massive 
than $M_{\rm h}^{\rm min}$ (e.g. \cite{ouchi2010}),
%
%
because LAEs should meet two physical conditions: 
i) LAEs should be active star-forming galaxies producing Ly$\alpha$ photons
and 
ii) LAE's Ly$\alpha$ photons should largely escape from the ISM.
In other words, not all of galaxies in a dark-matter halo
can be LAEs.
%
%
We evaluate the duty cycle of LAEs (i.e. $f_{\rm duty}^{\rm LAE}$) 
%
%
that is defined by the probability that a dark-matter halo beyond the minimum-halo mass hosts an LAE(s).
%
%
With $f_{\rm duty}$, one can calculate the galaxy number density,
\begin{equation}
n_{\rm g}  = \int_{M_{\rm h}^{\rm min}}^{\infty}\ f_{\rm duty}^{\rm LAE}\ N_{\rm g}\ n\ dM.
\label{eq:galaxy_number_density}
\end{equation}
We derive $f_{\rm duty}^{\rm LAE}$ with the best-fit HOD parameters,
assuming that $f_{\rm duty}^{\rm LAE}$ does not depend on the dark-matter halo mass.
The observational measurements of $n_{\rm g}$ are taken from the HSC SILVERRUSH paper of \citet{konno2017}.
We thus obtain 
$f_{\rm duty}^{\rm LAE}=7.7^{+53.0}_{-0.75} \times 10^{-4}$ and $0.9^{+13.5}_{-0.9} \times 10^{-4}$
for our LAEs at $z=5.7$ and $6.6$, respectively, that are
listed in Table \ref{tab:clustering_hod}.
Although $f_{\rm duty}^{\rm LAE}$ is very poorly constrained
with the large uncertainties, we find that the duty cycle of LAEs is $\sim 1$\% or less
at $z=5.7$ and $6.6$.
%
%
These small duty cycle values for $z=6-7$ LAEs are consistent with the previous
estimates by the similar technique \citep{ouchi2010}. The small duty cycle values
indicate that dark-matter halos hosting LAEs are rare, and that
majority of LAEs are populated in abundant low-mass dark matter halos
with a mass close to the minimum-halo mass.

\subsubsection{Galaxy Formation and LAEs}
\label{sec:galaxy_formation_LAEs}

Figure \ref{fig:redshift_bias_LBGLAE} presents the evolutionary
tracks of dark-matter halos for the galaxy-conserving model
and the constant-mass model.
For the galaxy-conserving models, we assume that the gravity
drives the motion of galaxies under the condition of no mergers.
The bias evolution in the galaxy-conserving model \citep{fry1996} 
is described as
%
%
%
\begin{equation}
b_{\rm g}=1+(b_{\rm g}^0-1)/D(z),
\label{eq:galaxy_conserving}
\end{equation}
where $D(z)$ and $b_{\rm g}^0$ are the growth factor and the bias at $z=0$,
respectively. If one assumes the galaxy-conserving evolution,
our LAEs at $z=5.7$ and $6.6$ evolve into the present-galaxies with
$b_{\rm g}^0\sim 1.6-1.7$. 
%
%
Galaxies with $b_{\rm g}^0\sim 1.6-1.7$
are massive bright $\sim 6 L^*$ galaxies today \citep{zehavi2005}. The galaxy-conserving evolution
suggests that our LAEs at $z=5.7$ and $6.6$ are the progenitors of 
very massive bright galaxies in the present-day universe.

\section{Summary}
\label{sec:summary}

We investigate clustering of Ly$\alpha$ emitters (LAEs) at $z=5.7$ and $6.6$
with the early data of the Hyper Suprime-Cam (HSC) Subaru Strategic Program 
(first-data release shown in \cite{aihara2017}), and introduce the program of
Systematic Identification of LAEs for Visible Exploration and Reionization Research 
Using Subaru HSC (SILVERRUSH). From the early data,
we obtain 
%
%
the unprecedentedly large samples of 1,832
%
%
$\gtrsim L^*$ LAEs at $z=6-7$ over 
the total area of $14-21$ deg$^2$ that is about an order of magnitude 
larger than the previous $z=6-7$ LAE clustering studies. Based on the LAE clustering
measurements, we study cosmic reionization and galaxy formation, comparing theoretical models.
In this study, there are two major results that are summarized below.

1. We calculate angular correlation functions of the $z=6-7$ LAEs.
The correlation lengths are estimated to be $r_0=3.01^{+0.35}_{-0.35}$ and $2.66^{+0.49}_{-0.70}$ $h^{-1}_{100}$ Mpc
for the $\gtrsim L^*$ LAEs at $z=5.7$ and $6.6$, respectively. The average of 
the large-scale bias value is $b_{\rm avg}=4.13\pm 0.17$ ($4.54\pm 0.63$) at $z=5.7$ ($z=6.6$) for the LAEs. 
Because Ly$\alpha$ photons emitted from LAEs in ionized bubbles can selectively escape from the partly 
neutral IGM at the EoR, observed LAEs at $z=6.6$ should have clustering signals stronger than the intrinsic
clustering. Based on this physical picture,
we obtain the constraint of 
%
%
%
$x_{\rm HI}=0.15^{+0.15}_{-0.15}$ 
%
%
at $z=6.6$ by the comparisons between our clustering measurements and
two independent theoretical models.

2. We study the $\gtrsim L^*$ LAE clustering by the halo occupation distribution (HOD) modeling.
(Here, for the LAE clustering at the EoR of $z=6.6$, we assume
that the LAE clustering is not largely impacted by the cosmic reionization.)
The best-fit models indicate that the $\gtrsim L^*$ LAEs are hosted by the dark-matter halos
with average masses of $\log (\left < M_{\rm h} \right >/M_\odot) =11.1^{+0.2}_{-0.4}$
and $10.8^{+0.3}_{-0.5}$ at $z=5.7$ and $6.6$, respectively.
Comparing the number densities of observed LAEs and those suggested from the HOD modeling,
we find that dark-matter halos of only 1 \% (or less) down to the minimum-halo mass
can host the $\gtrsim L^*$ LAEs. With the standard structure formation models,
the bias evolution of the dark-matter halos 
indicates that the $\gtrsim L^*$ LAEs at $z=6-7$ are progenitors of 
massive $\sim 6 L^*$ galaxies in the present-day universe.

\begin{ack}
We are grateful to useful discussion
with Mark Dijikstra, Richard Ellis, Andrea Ferrara,
Martin Haehnelt, Alex Hagen, Koki Kakiichi, Andrei Mesinger, 
Naveen Reddy, and Zheng Zheng.
%
%
We acknowledge Jirong Mao and Anne Hutter
for their comments.
%
%
%
The Hyper Suprime-Cam (HSC) collaboration includes the astronomical communities of Japan and Taiwan, and Princeton University. The HSC instrumentation and software were developed by the National Astronomical Observatory of Japan (NAOJ), the Kavli Institute for the Physics and Mathematics of the Universe (Kavli IPMU), the University of Tokyo, the High Energy Accelerator Research Organization (KEK), the Academia Sinica Institute for Astronomy and Astrophysics in Taiwan (ASIAA), and Princeton University. Funding was contributed by the FIRST program from Japanese Cabinet Office, the Ministry of Education, Culture, Sports, Science and Technology (MEXT), the Japan Society for the Promotion of Science (JSPS), Japan Science and Technology Agency (JST), the Toray Science Foundation, NAOJ, Kavli IPMU, KEK, ASIAA, and Princeton University. 
The NB816 filter was supported by Ehime University (PI: Y. Taniguchi).
The NB921 filter was supported by KAKENHI (23244025) Grant-in-Aid for Scientific 
Research (A) through the Japan Society for the Promotion of Science (PI: M. Ouchi). 
This paper makes use of software developed for the Large Synoptic Survey Telescope. We thank the LSST Project for making their code available as free software at  http://dm.lsst.org.
The Pan-STARRS1 Surveys (PS1) have been made possible through contributions of the Institute for Astronomy, the University of Hawaii, the Pan-STARRS Project Office, the Max-Planck Society and its participating institutes, the Max Planck Institute for Astronomy, Heidelberg and the Max Planck Institute for Extraterrestrial Physics, Garching, The Johns Hopkins University, Durham University, the University of Edinburgh, Queen’s University Belfast, the Harvard-Smithsonian Center for Astrophysics, the Las Cumbres Observatory Global Telescope Network Incorporated, the National Central University of Taiwan, the Space Telescope Science Institute, the National Aeronautics and Space Administration under Grant No. NNX08AR22G issued through the Planetary Science Division of the NASA Science Mission Directorate, the National Science Foundation under Grant No. AST-1238877, the University of Maryland, and Eotvos Lorand University (ELTE) and the Los Alamos National Laboratory.
This work is supported by World Premier International Research 
Center Initiative (WPI Initiative), MEXT, Japan, and 
KAKENHI (15H02064) Grant-in-Aid for Scientific Research (A) 
through Japan Society for the Promotion of Science.
Based on data collected at the Subaru Telescope and retrieved from the HSC data archive system, which is operated by Subaru Telescope and Astronomy Data Center at National Astronomical Observatory of Japan.
%
\end{ack}

%
%


\end{document}